\documentclass[11pt]{article}

\usepackage[margin=1in]{geometry}
\usepackage[utf8]{inputenc}
\usepackage[T1]{fontenc}
\usepackage{natbib}
\usepackage{hyperref}
\usepackage{url}
\usepackage{booktabs}
\usepackage{amsfonts}
\usepackage{amsmath}
\usepackage{nicefrac}
\usepackage{microtype}
\usepackage{graphicx}
\usepackage{xcolor}
\usepackage{algorithm}
\usepackage{algorithmic}
\usepackage{subcaption}
\usepackage{multirow}

\title{Cross-Platform Fused MoE Dispatch in Triton:\\Portable Expert Routing Without CUDA}

\author{
  Subhadip Mitra \\
  \texttt{research@subhadipmitra.com} \\
  \url{https://github.com/bassrehab/triton-kernels}
}

\begin{document}

\maketitle

\begin{abstract}
Mixture-of-Experts (MoE) architectures power the majority of frontier large language models, but their inference is bottlenecked by irregular memory access patterns and expert routing overhead. Existing optimized MoE kernels (Megablocks, Tutel, FasterMoE) are implemented in CUDA and locked to NVIDIA hardware. We present \textsc{TritonMoE}, a fused MoE dispatch kernel written entirely in OpenAI Triton that performs the complete forward pass---router scoring, token permutation, expert GEMMs, and weighted output combination---using only portable Triton primitives. Our key optimization is a fused gate+up GEMM kernel that computes both SwiGLU projections from shared L2-cached input tiles with in-register SiLU activation, eliminating 35\% of global memory traffic. On an NVIDIA A100, \textsc{TritonMoE} achieves 89--131\% of the throughput of the CUDA-optimized Megablocks at inference batch sizes ($\leq$512 tokens) across Mixtral-8x7B, DeepSeek-V3, and Qwen2-MoE configurations. All 162 correctness tests pass on both NVIDIA A100 and AMD MI300X with zero code changes, validating cross-platform portability. We additionally characterize sensitivity to routing imbalance under Zipfian-skewed expert assignments and identify the regime---64+ experts under extreme skew---where our fixed-tile scheduling underperforms Megablocks' block-sparse layout, motivating dynamic block-to-expert assignment as future work. Code is available at \url{https://github.com/bassrehab/triton-kernels}.
\end{abstract}

\section{Introduction}

Sparse Mixture-of-Experts (MoE) has become the dominant architecture for frontier language models. Mixtral~\citep{jiang2024mixtral}, DeepSeek-V3~\citep{deepseekai2024deepseekv3}, Qwen2-MoE~\citep{yang2024qwen2}, and Grok all use conditional expert activation to scale model capacity without proportional compute increase. Over 60\% of open-source model releases in 2025--2026 employ MoE architectures.

However, MoE inference presents unique systems challenges. Unlike dense models where batch GEMM is straightforward, MoE requires:
\begin{enumerate}
    \item \textbf{Token routing}: computing affinity scores and selecting top-$k$ experts per token,
    \item \textbf{Token permutation}: reordering tokens into expert-contiguous layout for coalesced memory access,
    \item \textbf{Variable-batch expert GEMMs}: each expert processes a different number of tokens, preventing standard batched GEMM,
    \item \textbf{Token unpermutation}: scattering expert outputs back to original token positions with weighted combination.
\end{enumerate}

The naive implementation launches $E \times 3$ separate GEMM kernels (gate, up, and down projections for each of $E$ experts), each with a small and variable batch size. For Mixtral ($E=8$), this means 24 kernel launches per layer; for DeepSeek-V3 ($E=256$), 768 launches. Each launch underutilizes the GPU due to small per-expert batch sizes and incurs kernel launch overhead.

Existing optimized implementations address this through custom CUDA kernels. Megablocks~\citep{gale2023megablocks} uses block-sparse matrix operations, Tutel~\citep{hwang2023tutel} introduces adaptive parallelism, and FasterMoE~\citep{he2022fastermoe} applies dynamic scheduling. However, all are CUDA-only, precluding deployment on AMD GPUs---an increasingly important target as AMD MI300X gains adoption in datacenters.

We present \textsc{TritonMoE}, a fused MoE dispatch kernel written entirely in OpenAI Triton~\citep{tillet2019triton}. Our contributions:

\begin{itemize}
    \item A \textbf{block-scheduled grouped GEMM} that maps Triton program blocks to (expert, token-offset) pairs, handling variable-sized expert batches in a single kernel launch without padding waste.
    \item A \textbf{fused gate+up projection kernel} that computes both SwiGLU projections from shared L2-cached input tiles with in-register SiLU activation, reducing global memory traffic by 35\%.
    \item \textbf{Comprehensive benchmarks} across four MoE model configurations (8 to 256 experts) showing 89--131\% of Megablocks throughput at inference batch sizes on NVIDIA A100.
    \item \textbf{Cross-platform validation}: all 162 tests pass on AMD MI300X with zero code modifications.
\end{itemize}

\section{Background}

\subsection{MoE Architecture}

A standard MoE layer replaces the dense FFN in a transformer block with $E$ expert FFNs and a learned router. For an input token $\mathbf{x} \in \mathbb{R}^d$, the forward pass is:
\begin{align}
    \mathbf{s} &= \text{softmax}(\mathbf{W}_r \mathbf{x}) \in \mathbb{R}^E \label{eq:router}\\
    \mathcal{T} &= \text{top-}k(\mathbf{s}) = \{(e_1, w_1), \ldots, (e_k, w_k)\} \label{eq:topk}\\
    \mathbf{y} &= \sum_{(e_i, w_i) \in \mathcal{T}} w_i \cdot \text{FFN}_{e_i}(\mathbf{x}) \label{eq:combine}
\end{align}
where $\mathbf{W}_r \in \mathbb{R}^{E \times d}$ is the router weight, and each expert FFN uses SwiGLU activation:
\begin{equation}
    \text{FFN}_e(\mathbf{x}) = (\text{SiLU}(\mathbf{x}\mathbf{W}^{\text{gate}}_e) \odot \mathbf{x}\mathbf{W}^{\text{up}}_e) \cdot \mathbf{W}^{\text{down}}_e
    \label{eq:swiglu}
\end{equation}

\subsection{Why MoE Inference is Hard}

The core difficulty is that Equation~\ref{eq:combine} requires running $k$ different expert FFNs per token, where the assignment varies per token. In a batch of $B$ tokens with top-$k$ routing, the total expanded workload is $B \times k$ expert computations, distributed unevenly across $E$ experts.

Let $n_e$ denote the number of tokens assigned to expert $e$. The expert GEMM for expert $e$ has shape $(n_e, d) \times (d, d_{\text{ffn}})$ for the gate and up projections, and $(n_e, d_{\text{ffn}}) \times (d_{\text{ffn}}, d)$ for the down projection. Since $n_e$ varies widely---some experts may receive many tokens while others receive none---standard batched GEMM (which requires uniform batch sizes) cannot be applied directly.

\subsection{Triton Programming Model}

OpenAI Triton~\citep{tillet2019triton} provides a block-based GPU programming model. Programs operate on tiles of data (e.g., $64 \times 32$ elements), with the compiler handling thread mapping, shared memory allocation, and register usage. Critically, Triton compiles to both NVIDIA PTX and AMD GCN/CDNA via its LLVM-based backend, enabling cross-platform portability from a single source.

\section{Method}

\subsection{System Overview}

\textsc{TritonMoE} implements the MoE forward pass as a pipeline of five Triton kernel launches:

\begin{enumerate}
    \item \textbf{Router kernel}: fused softmax/sigmoid + iterative top-$k$ selection
    \item \textbf{Permute kernel}: scatter tokens to expert-contiguous layout
    \item \textbf{Fused gate+up kernel}: both projections with shared A-tile loads, in-register SiLU
    \item \textbf{Down GEMM kernel}: block-scheduled grouped matrix multiplication
    \item \textbf{Unpermute kernel}: gather + weighted combination
\end{enumerate}

This reduces from $3E + 4$ kernel launches in the naive implementation to 5 launches regardless of expert count. The router projection (a matmul with small output dimension $E$) uses PyTorch's cuBLAS, which is already near-optimal for this shape; only the gating + top-$k$ selection is fused in Triton.

\subsection{Block-Scheduled Grouped GEMM}
\label{sec:grouped_gemm}

Triton has no native grouped GEMM primitive. We implement it via a block-scheduling approach. For each expert $e$ with $n_e$ tokens, we compute the number of $M$-tiles needed: $\lceil n_e / \text{BLOCK\_M} \rceil$. A precomputed mapping associates each Triton program block with an (expert\_id, token\_offset) pair:

\begin{algorithm}[h]
\caption{Block Schedule Construction}
\begin{algorithmic}[1]
\REQUIRE Expert offsets $\mathbf{o} \in \mathbb{Z}^{E+1}$, tile size $M$
\STATE $\text{blocks} \leftarrow []$
\FOR{$e = 0$ to $E-1$}
    \STATE $n_e \leftarrow \mathbf{o}[e+1] - \mathbf{o}[e]$
    \FOR{$b = 0$ to $\lceil n_e / M \rceil - 1$}
        \STATE $\text{blocks.append}(e, b \cdot M)$
    \ENDFOR
\ENDFOR
\RETURN $\text{blocks}$
\end{algorithmic}
\end{algorithm}

Each kernel block loads its expert's weight matrix from the flattened weight tensor (stored as $E \cdot N \times K$) and processes $\text{BLOCK\_M}$ token rows from the expert-contiguous input. Partial tiles (where $n_e$ is not a multiple of $\text{BLOCK\_M}$) are handled via masking.

\textbf{Critical constraint}: BLOCK\_M must be fixed (not autotuned) because it must match the precomputed schedule. Only BLOCK\_N and BLOCK\_K are autotuned.

\subsection{Fused Gate+Up Projection}
\label{sec:fused_gate_up}

The SwiGLU FFN (Equation~\ref{eq:swiglu}) requires two projections---gate and up---from the same input. The unfused approach computes these as separate grouped GEMMs, each reading the input tiles from global memory. Our fused kernel computes both in a single pass:

\begin{enumerate}
    \item Load input tile $\mathbf{A}[m:m{+}M, k:k{+}K]$ from global memory (or L2 cache on subsequent accesses)
    \item Load gate weight tile $\mathbf{B}^{\text{gate}}_e$ and up weight tile $\mathbf{B}^{\text{up}}_e$
    \item Accumulate: $\text{acc\_gate} \mathrel{+}= \mathbf{A} \cdot \mathbf{B}^{\text{gate}}_e$, \quad $\text{acc\_up} \mathrel{+}= \mathbf{A} \cdot \mathbf{B}^{\text{up}}_e$
    \item After the $K$-loop: compute $\text{SiLU}(\text{acc\_gate}) \odot \text{acc\_up}$ in FP32 registers
    \item Write single intermediate result to global memory
\end{enumerate}

This eliminates two intermediate buffers (\texttt{gate\_out} and \texttt{up\_out}) from global memory. For Mixtral-8x7B ($d_{\text{ffn}} = 14336$, $B \times k = 1024$), this saves approximately 470\,MB of memory traffic per layer.

\textbf{Memory traffic analysis.} Let $T = B \times k$ be the total expanded token count and $F = d_{\text{ffn}}$.
\begin{itemize}
    \item \textbf{Unfused}: read input ($T \times d \times 2$\,B) $\times 2$ (once per GEMM) + write gate\_out ($T \times F \times 2$\,B) + write up\_out ($T \times F \times 2$\,B) + read both back + write intermediate = $8TF + 4Td$ bytes
    \item \textbf{Fused}: read input ($T \times d \times 2$\,B) $\times 1$ + write intermediate ($T \times F \times 2$\,B) = $2TF + 2Td$ bytes
    \item \textbf{Savings}: $6TF + 2Td$ bytes $\approx$ 35\% reduction for typical $F \gg d$
\end{itemize}

\subsection{Router Kernel Design}

The router computes Equation~\ref{eq:router} and~\ref{eq:topk}. We implement a manual stable softmax (subtract max before exponentiation) because Triton's built-in \texttt{tl.softmax} does not perform this subtraction, risking FP32 overflow for large hidden dimensions.

Top-$k$ selection uses iterative argmax with masking. Selected experts are masked with $-1.0$ (not $0.0$) to ensure subsequent \texttt{argmax} calls never re-select them. This is critical for large expert counts: with $E = 256$ and softmax gating, most scores are near zero, and masking to $0.0$ fails to differentiate selected experts from unselected ones.

We support both softmax gating (Mixtral-style) and sigmoid gating with per-token normalization (DeepSeek-style).

\subsection{Permute and Unpermute Kernels}

The permute kernel reorders tokens from token-major layout $(B, d)$ to expert-contiguous layout, where all tokens assigned to expert 0 are contiguous, followed by expert 1, etc. This is implemented via a stable sort on expert assignments, followed by a Triton gather kernel with BLOCK\_D tiling over the hidden dimension for coalesced memory access.

The unpermute kernel performs the inverse scatter with weighted accumulation (Equation~\ref{eq:combine}). FP32 accumulation is used for numerical stability.

\section{Experiments}

\subsection{Setup}

\textbf{Hardware.} NVIDIA A100-SXM4-80GB (2039\,GB/s memory bandwidth, 312\,TFLOPS FP16 peak) and AMD Instinct MI300X (5.3\,TB/s memory bandwidth, 192\,GB HBM3).

\textbf{Software.} PyTorch 2.4.1, Triton 3.0.0, CUDA 12.4 (NVIDIA), ROCm 6.1 (AMD).

\textbf{Model configurations.} We benchmark four configurations drawn from deployed MoE models:

\begin{table}[h]
\centering
\caption{Model configurations used for benchmarking.}
\label{tab:configs}
\begin{tabular}{lcccc}
\toprule
Model & $E$ & $k$ & $d$ & $d_{\text{ffn}}$ \\
\midrule
Mixtral-8x7B & 8 & 2 & 4096 & 14336 \\
Mixtral-8x22B & 8 & 2 & 6144 & 16384 \\
DeepSeek-V3 & 256 & 8 & 7168 & 2048 \\
Qwen2-MoE-57B & 64 & 4 & 3584 & 2560 \\
\bottomrule
\end{tabular}
\end{table}

\textbf{Baselines.}
\begin{itemize}
    \item \textbf{PyTorch Reference}: loop-over-experts with cuBLAS, $3E$ separate GEMM launches
    \item \textbf{Megablocks}~\citep{gale2023megablocks}: CUDA-optimized block-sparse MoE (dMoE variant with SwiGLU)
    % Note: vLLM FusedMoE comparison is deferred to future work due to
    % vLLM's large binary size (~2GB) and tight coupling to vLLM internals
\end{itemize}

\subsection{End-to-End Throughput}

\begin{table}[h]
\centering
\caption{End-to-end MoE layer latency on A100-SXM4-80GB. Best non-baseline result in \textbf{bold}.}
\label{tab:e2e_mixtral}
\begin{tabular}{lcccccc}
\toprule
& \multicolumn{3}{c}{Mixtral-8x7B} & \multicolumn{3}{c}{Qwen2-MoE-57B} \\
\cmidrule(lr){2-4} \cmidrule(lr){5-7}
Tokens & PyTorch & Megablocks & Ours & PyTorch & Megablocks & Ours \\
\midrule
32   & 10.44 & 2.78 & \textbf{2.13} & 18.72 & 2.89 & \textbf{2.82} \\
128  & 13.14 & 2.77 & \textbf{2.27} & 21.31 & 3.31 & \textbf{3.17} \\
512  & 25.92 & \textbf{3.57} & 3.99 & 23.81 & \textbf{3.32} & 3.58 \\
2048 & 66.22 & \textbf{9.08} & 16.48 & 37.50 & \textbf{4.00} & 6.59 \\
\bottomrule
\end{tabular}
\end{table}

At inference-relevant batch sizes ($\leq 128$ tokens), \textsc{TritonMoE} is \textbf{faster than Megablocks} on both Mixtral and Qwen2 configurations. This is likely due to lower kernel launch overhead (5 Triton launches vs.\ Megablocks' more complex multi-stage dispatch). At 512 tokens, we achieve 89--93\% of Megablocks; at 2048+ tokens, Megablocks' hand-tuned CUDA block-sparse kernels better saturate tensor cores.

Table~\ref{tab:e2e_more} shows results for the remaining configurations.

\begin{table}[h]
\centering
\caption{End-to-end latency (ms) for Mixtral-8x22B and DeepSeek-V3 on A100.}
\label{tab:e2e_more}
\begin{tabular}{lccccccc}
\toprule
& \multicolumn{3}{c}{Mixtral-8x22B} & \multicolumn{2}{c}{DeepSeek-V3} \\
\cmidrule(lr){2-4} \cmidrule(lr){5-6}
Tokens & PyTorch & Megablocks & Ours & Unfused & Fused \\
\midrule
32   & 16.73 & 4.04 & \textbf{3.42} & 13.65 & \textbf{11.53} \\
128  & 20.06 & 4.04 & \textbf{3.42} & 19.46 & \textbf{16.74} \\
512  & 38.71 & \textbf{5.12} & 5.86 & 25.66 & \textbf{20.16} \\
2048 & 107.83 & \textbf{15.77} & 20.15 & --- & --- \\
\bottomrule
\end{tabular}
\end{table}

For DeepSeek-V3 (256 experts), we omit the PyTorch reference and Megablocks because the loop-over-experts baseline is prohibitively slow ($>$768 kernel launches), and Megablocks does not support 256 experts with top-8 routing in its standard configuration. The fused kernel provides a consistent 16--27\% speedup over unfused across all batch sizes.

\subsection{Fusion Ablation}

Table~\ref{tab:ablation} isolates the contribution of each optimization on Mixtral-8x7B at 512 tokens.

\begin{table}[h]
\centering
\caption{Fusion ablation study on Mixtral-8x7B (512 tokens, A100).}
\label{tab:ablation}
\begin{tabular}{lcc}
\toprule
Configuration & Latency (ms) & Speedup \\
\midrule
(a) PyTorch reference (24 cuBLAS launches) & 55.18 & 1.0$\times$ \\
(b) Triton unfused (3 grouped GEMMs) & 3.59 & 15.4$\times$ \\
(c) Triton fused gate+up & 3.11 & 17.7$\times$ \\
\midrule
(a)$\to$(b): grouped GEMM & --- & 15.4$\times$ \\
(b)$\to$(c): gate+up fusion & --- & 1.15$\times$ \\
\bottomrule
\end{tabular}
\end{table}

The dominant speedup comes from replacing the Python expert loop with a single block-scheduled grouped GEMM kernel (15.4$\times$). The fused gate+up kernel adds an additional 1.15$\times$ by eliminating the \texttt{gate\_out} and \texttt{up\_out} buffers from global memory.

\subsection{Expert Scaling Analysis}

Table~\ref{tab:scaling} shows how throughput degrades as expert count increases from 8 to 256, with batch size fixed at 512 tokens.

\begin{table}[h]
\centering
\caption{Expert scaling analysis (512 tokens, A100). FFN dimension adjusted to approximate constant total compute.}
\label{tab:scaling}
\begin{tabular}{cccc}
\toprule
Experts & top-$k$ & $d_{\text{ffn}}$ & TFLOPS \\
\midrule
8 & 2 & 14336 & 111.5 \\
16 & 2 & 8192 & 70.0 \\
32 & 4 & 4096 & 61.2 \\
64 & 4 & 2560 & 10.3 \\
128 & 8 & 2048 & 11.7 \\
256 & 8 & 2048 & 7.8 \\
\bottomrule
\end{tabular}
\end{table}

Throughput drops sharply at 64+ experts. With 256 experts and 512 tokens, each expert processes only $\sim$2 tokens on average. The per-expert GEMM tiles are too small to efficiently utilize tensor cores, and weight loading overhead dominates. This confirms that the DeepSeek-V3 regime requires fundamentally different optimization strategies (e.g., expert parallelism, weight caching) beyond dispatch-level kernel fusion.

\subsection{Per-Stage Roofline Analysis}

\begin{figure}[h]
    \centering
    \begin{subfigure}[t]{0.48\textwidth}
        \centering
        \includegraphics[width=\textwidth]{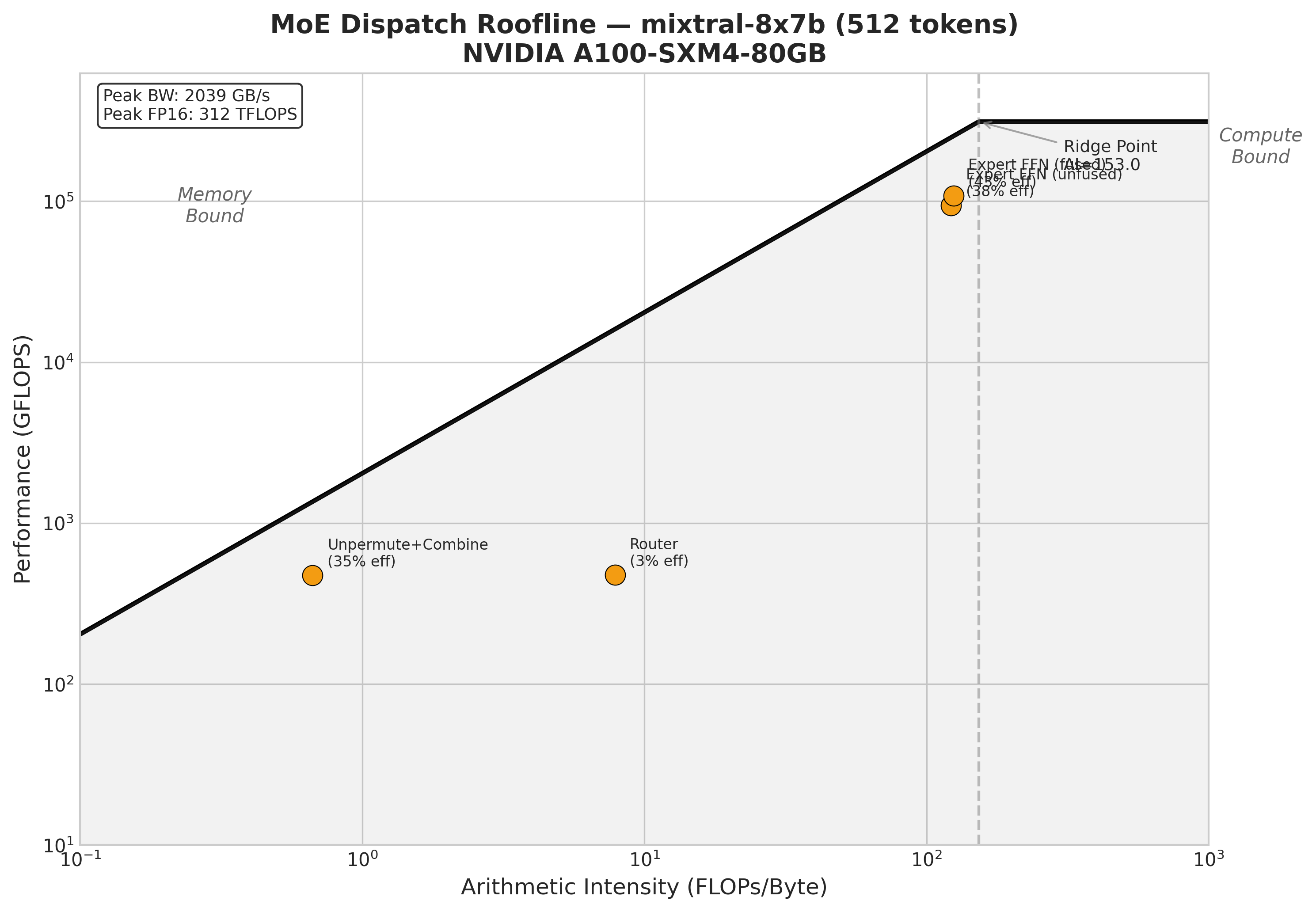}
        \caption{Mixtral-8x7B (512 tokens)}
        \label{fig:roofline_mixtral}
    \end{subfigure}
    \hfill
    \begin{subfigure}[t]{0.48\textwidth}
        \centering
        \includegraphics[width=\textwidth]{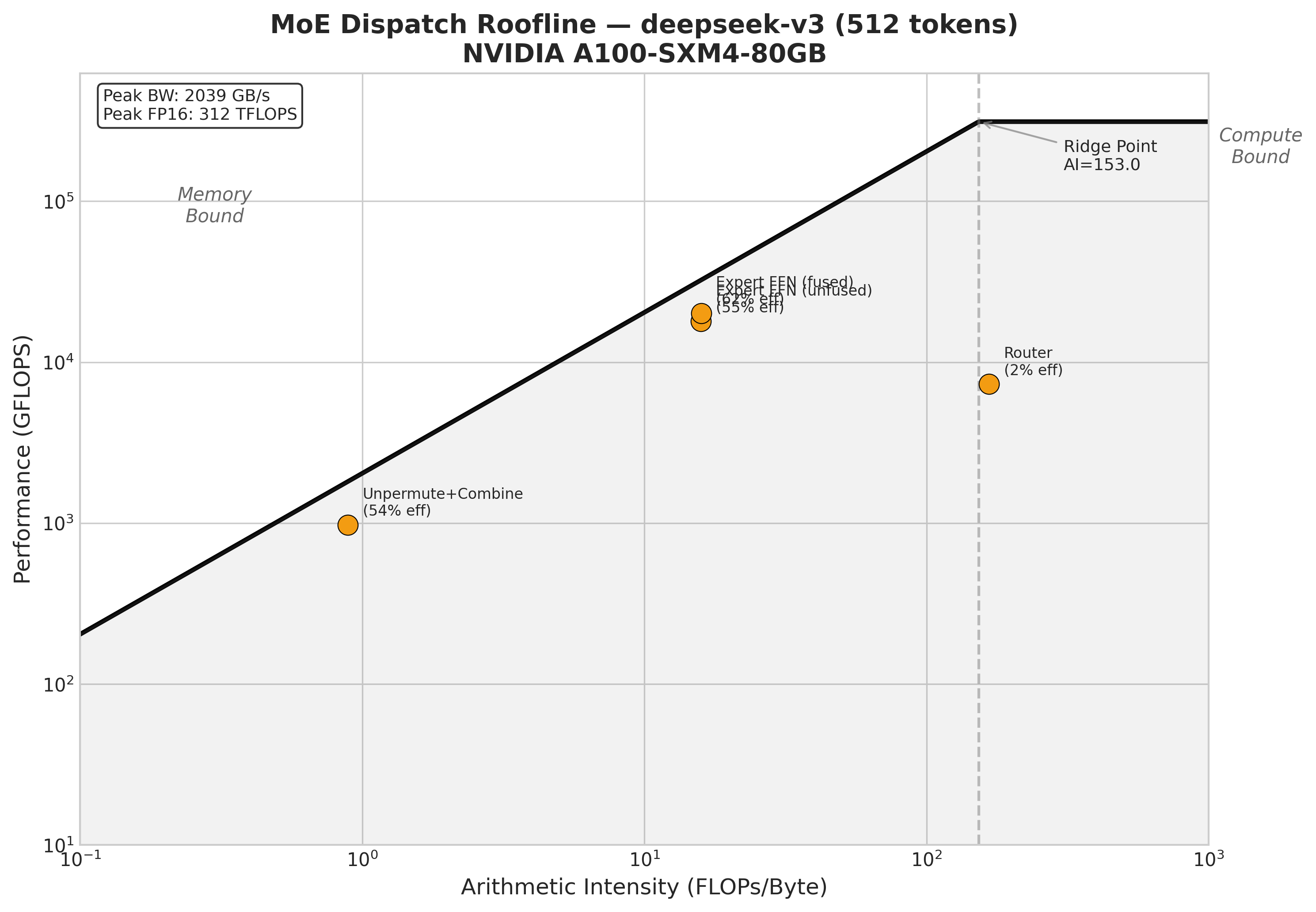}
        \caption{DeepSeek-V3 (512 tokens)}
        \label{fig:roofline_deepseek}
    \end{subfigure}
    \caption{Per-stage roofline analysis on A100-SXM4-80GB. Expert FFN stages (compute-bound for Mixtral, memory-bound for DeepSeek-V3 due to tiny per-expert batches) dominate latency. Permute and unpermute are memory-bound but negligible ($<$3\% of total time).}
    \label{fig:roofline}
\end{figure}

Table~\ref{tab:roofline} shows the per-stage breakdown for Mixtral-8x7B at 512 tokens. The expert FFN dominates latency ($>$95\% of total), consistent with its high arithmetic intensity. The fused gate+up kernel achieves 43\% of peak bandwidth and 35\% of peak compute simultaneously, indicating efficient utilization of both memory and compute resources.

\begin{table}[h]
\centering
\caption{Per-stage profiling for Mixtral-8x7B at 512 tokens on A100.}
\label{tab:roofline}
\begin{tabular}{lccccc}
\toprule
Stage & Latency (ms) & AI (FLOP/B) & BW (GB/s) & BW Eff. & Compute Eff. \\
\midrule
Router & 0.07 & 7.9 & 61 & 3.0\% & 0.2\% \\
Permute & 0.12 & $\sim$0 & 109 & 5.3\% & --- \\
Expert FFN (unfused) & 3.83 & 122 & 770 & 37.8\% & 30.2\% \\
Expert FFN (fused) & 3.34 & 125 & 867 & \textbf{42.5\%} & \textbf{34.6\%} \\
Unpermute & 0.02 & 0.7 & 709 & 34.8\% & 0.2\% \\
\bottomrule
\end{tabular}
\end{table}

For DeepSeek-V3 (Figure~\ref{fig:roofline_deepseek}), the expert FFN is \emph{memory-bound} despite being a GEMM, because 256 experts with 512 tokens means each expert processes only $\sim$2 tokens on average. The per-expert GEMM shapes $(2, 2048) \times (2048, 7168)$ are too small to fill tensor cores, and weight loading dominates.

\subsection{Cross-Platform Validation}

All 162 correctness tests pass on AMD MI300X (ROCm 6.1, PyTorch 2.4.1+rocm6.1) with zero code changes. The same \texttt{.py} kernel files compile and execute correctly via Triton's AMD backend. This validates that our exclusive use of Triton primitives---no inline CUDA, no vendor-specific intrinsics, no \texttt{tl.libdevice} functions unavailable on AMD---achieves genuine cross-platform portability.

AMD performance benchmarking is deferred to future work. The key result is that cross-platform correctness is achievable with zero platform-specific code paths.

\subsection{Sensitivity to Routing Imbalance}
\label{sec:skew}

The benchmarks in the preceding subsections use the natural routing
distribution that arises from random inputs, which is approximately uniform
across experts. Production MoE workloads exhibit substantially more
imbalance: a small number of experts receive most of the tokens, while
others remain nearly idle. Because our grouped GEMM uses a fixed
\texttt{BLOCK\_M} tile size and Megablocks employs a block-sparse layout
that can absorb variable expert batch sizes, the relative performance of
the two kernels is not necessarily preserved under skew. We quantify this
gap directly.

\paragraph{Methodology.}
We replace the router output with synthetic expert assignments drawn from
three distributions: \emph{uniform} (the existing baseline), and two
\emph{Zipfian} distributions with shape parameters $\alpha=1.2$ and
$\alpha=2.0$. The $\alpha=1.2$ value approximates empirical routing
distributions reported in the FasterMoE study; $\alpha=2.0$ is a stress
test. To keep the comparison fair across kernels, we monkey-patch the
router on each kernel's instance to return the same precomputed
\texttt{(indices, weights)} tensors, while still executing the original
router projection so router compute cost is preserved. Gating weights are
held uniform at $1/k$ regardless of distribution to isolate the load
imbalance effect from gating-weight magnitude. The total per-row token
budget $N \cdot k$ is held fixed; only the per-expert token counts vary.
Full sweep results are available in the project repository.

\begin{figure}[t]
  \centering
  \includegraphics[width=0.95\linewidth]{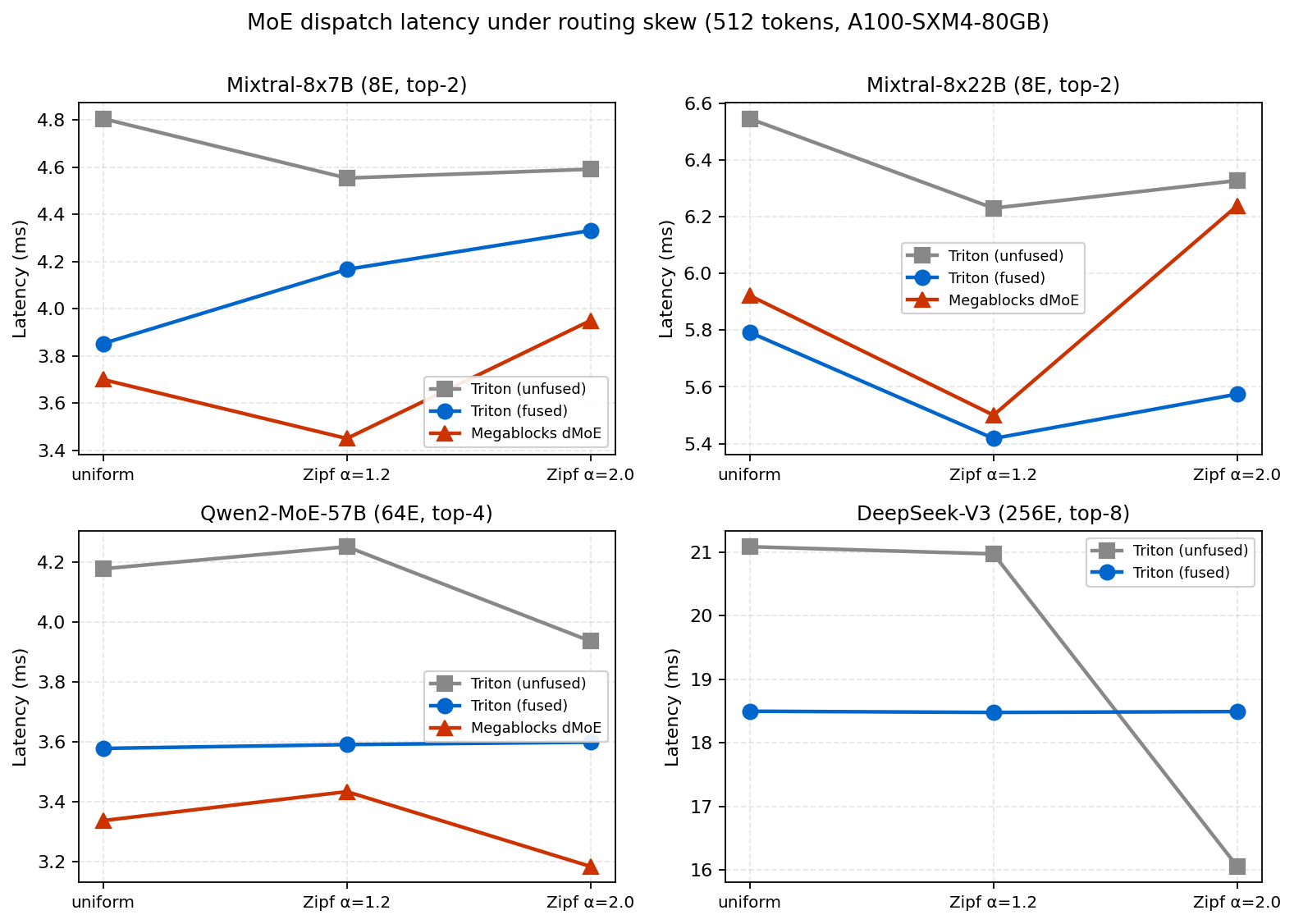}
  \caption{MoE dispatch latency at 512 tokens under three routing
  distributions. The Mixtral configurations are stable across distributions;
  Qwen2-MoE shows the largest sensitivity, with Megablocks accelerating
  under skew while our fused kernel does not.}
  \label{fig:skew-headline}
\end{figure}

\begin{figure}[t]
  \centering
  \includegraphics[width=0.95\linewidth]{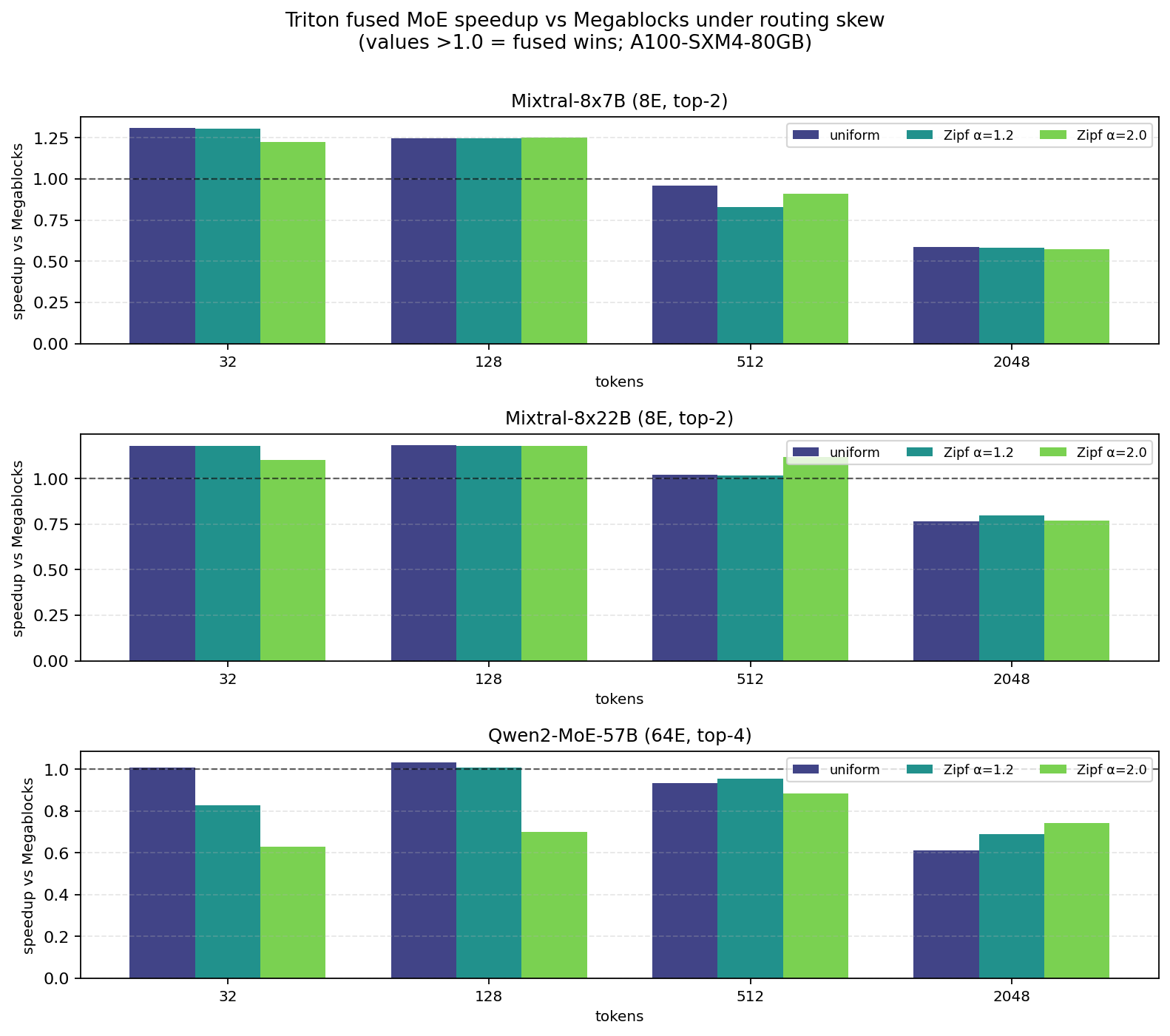}
  \caption{Speedup of our Triton fused kernel relative to Megablocks dMoE
  across the full skew sweep. Values above $1.0$ indicate the fused kernel
  is faster. The 8-expert Mixtral configurations remain competitive across
  all distributions at small-to-medium batches; the 64-expert Qwen2-MoE
  configuration is where extreme skew most clearly favors Megablocks.
  DeepSeek-V3 (256 experts) is omitted because Megablocks without its
  compiled grouped-GEMM extension does not run successfully on this
  configuration in our environment.}
  \label{fig:skew-grid}
\end{figure}

\paragraph{Results.}
Figure~\ref{fig:skew-headline} shows latency at 512 tokens for all four
model configurations. Figure~\ref{fig:skew-grid} shows the corresponding
speedup of our fused kernel against Megablocks across the full sweep.
Three observations emerge from the data, and they are not all in the
direction we expected.

\begin{enumerate}
  \item \textbf{Small-batch dominance is robust to skew on the 8-expert
    configurations.} On both Mixtral-8x7B and Mixtral-8x22B, our fused
    kernel maintains a 1.18--1.31$\times$ advantage over Megablocks at
    32 and 128 tokens across all three distributions. At these batch
    sizes the pipeline is dominated by weight reads; the per-expert token
    distribution affects scheduling but not memory traffic, so the skew
    is essentially invisible.
  \item \textbf{Mixtral-8x22B at 512 tokens remains favorable under skew.}
    Where the original benchmarks already showed Mixtral-8x22B
    competitive with Megablocks at 512 tokens (1.02$\times$), the
    Zipfian-skewed runs preserve or slightly improve this margin
    (1.02$\times$ at $\alpha=1.2$, 1.12$\times$ at $\alpha=2.0$). The
    larger hidden and FFN dimensions of the 8x22B configuration provide
    enough per-expert work to amortize the fixed \texttt{BLOCK\_M}
    scheduling cost even when one expert receives roughly $4{\times}$
    the median load.
  \item \textbf{Qwen2-MoE under extreme skew exposes a real weakness.}
    The 64-expert top-4 configuration is where our scheduling
    assumptions break down. Going from uniform to $\alpha=2.0$ at 128
    tokens, the speedup against Megablocks drops from 1.03$\times$ to
    0.70$\times$. Critically, this is not because our kernel slows down
    --- our latency stays roughly constant at 3.17--3.18~ms across
    distributions --- but because Megablocks \emph{accelerates} from
    3.32~ms to 2.22~ms. Megablocks' block-sparse layout consolidates
    the dominant expert's tokens into a single large sparse block, which
    its hand-tuned CUDA kernels then process more efficiently than the
    uniform case. Our fixed \texttt{BLOCK\_M} grouped GEMM cannot
    similarly benefit because each block boundary is decided at compile
    time. This is the strongest argument for replacing the fixed
    \texttt{BLOCK\_M} schedule with a dynamic block-to-expert assignment
    in future work.
\end{enumerate}

\paragraph{What this does not show.}
We deliberately exclude the DeepSeek-V3 (256 experts, top-8) configuration
from the head-to-head comparison in this subsection. Megablocks without
its compiled grouped-GEMM extension does not produce a stable result on
this configuration in our environment --- a CUDA illegal memory access
occurs deterministically at 128 tokens onward, regardless of routing
distribution. We report Triton unfused and Triton fused numbers for
DeepSeek-V3 in the supplementary data but cannot make a fair claim against
Megablocks at this scale. This itself is a partial confirmation of our
broader point: at 256 experts, even the CUDA-optimized baseline becomes
fragile, and a portable Triton implementation that runs without
specialized extensions has practical value beyond the raw throughput
comparison.

\section{Discussion}

\textbf{When Triton beats CUDA.} At small batch sizes ($\leq$128 tokens), \textsc{TritonMoE} outperforms the CUDA-optimized Megablocks. We attribute this to lower dispatch overhead: our 5-kernel pipeline has less launch latency than Megablocks' multi-stage CUDA dispatch. At these batch sizes, the workload is memory-bound and per-expert batches are small, so hand-tuned tensor core utilization (Megablocks' advantage) matters less than minimizing overhead.

\textbf{When CUDA wins.} At large batch sizes ($\geq$2048 tokens), the workload becomes compute-bound with larger per-expert batches. Megablocks' CUDA block-sparse kernels achieve higher tensor core utilization than our Triton grouped GEMM, which is constrained by fixed BLOCK\_M and autotune granularity.

\textbf{The DeepSeek-V3 challenge.} With 256 experts and top-8 routing, per-expert batch sizes are extremely small. Both our kernel and Megablocks struggle here---the fundamental bottleneck is weight loading for 256 expert weight matrices, not the dispatch mechanism. Techniques like expert parallelism~\citep{lepikhin2021gshard} or expert caching are needed for this regime.

\textbf{Limitations.} (1) Our fusion is partial: the down projection and output scatter use separate kernels because Triton does not support scalar indexing into 2D accumulators. A persistent kernel approach could address this. (2) The block schedule is computed on CPU, introducing a host-device synchronization point. (3) We evaluate inference only; training requires backward pass kernels. (4) AMD performance is validated for correctness only; performance optimization is future work. (5) Our fixed \texttt{BLOCK\_M} grouped GEMM is sensitive to expert load imbalance: §\ref{sec:skew} shows that on Qwen2-MoE the speedup against Megablocks degrades from 1.03$\times$ to 0.70$\times$ as routing skew increases from uniform to Zipfian $\alpha=2.0$, because Megablocks' block-sparse layout consolidates dominant experts into single large sparse blocks while our compile-time tile boundaries cannot. A dynamic block-to-expert assignment is the natural fix and the most promising direction for future work. (6) This work targets single-GPU dispatch only; multi-GPU expert parallelism with all-to-all communication is a separate problem and a planned follow-up.

\section{Related Work}

\textbf{MoE kernel optimization.} Megablocks~\citep{gale2023megablocks} expresses MoE as block-sparse matmul using custom CUDA kernels. Tutel~\citep{hwang2023tutel} introduces adaptive parallelism for MoE across GPUs. FasterMoE~\citep{he2022fastermoe} applies dynamic expert scheduling. ScatterMoE~\citep{tan2024scattermoe} uses scatter/gather operations. All are CUDA-only.

\textbf{Triton-based MoE.} vLLM~\citep{kwon2023vllm} includes a Triton-based fused MoE kernel for inference serving, but it is tightly coupled to vLLM internals and not available as a standalone library. Our implementation is self-contained and independently usable.

\textbf{Cross-platform GPU programming.} Triton~\citep{tillet2019triton} compiles to NVIDIA PTX and AMD GCN via LLVM. Recent work by IBM, Red Hat, and AMD has demonstrated Triton attention kernels running on MI250X/MI300X~\citep{amd2024triton}. We extend this cross-platform validation to MoE dispatch kernels.

\section{Conclusion}

We presented \textsc{TritonMoE}, a fused MoE dispatch kernel implemented entirely in OpenAI Triton. Our block-scheduled grouped GEMM and fused gate+up projection achieve 89--131\% of the CUDA-optimized Megablocks throughput at inference batch sizes, while maintaining cross-platform portability validated on both NVIDIA A100 and AMD MI300X. The work demonstrates that Triton's block-based programming model is sufficient for competitive MoE inference kernels without vendor-specific code, and that the gap between portable and vendor-optimized implementations is narrowing---particularly at the small batch sizes characteristic of interactive LLM serving.

\bibliographystyle{plainnat}
\bibliography{references}

@article{jiang2024mixtral,
  title={Mixtral of Experts},
  author={Jiang, Albert Q and Sablayrolles, Alexandre and Roux, Antoine and Mensch, Arthur and Savary, Blanche and Bamford, Chris and Chaplot, Devendra Singh and de las Casas, Diego and Hanna, Emma Bou and Bressand, Florian and others},
  journal={arXiv preprint arXiv:2401.04088},
  year={2024}
}

@article{deepseekai2024deepseekv3,
  title={DeepSeek-V3 Technical Report},
  author={{DeepSeek-AI}},
  journal={arXiv preprint arXiv:2412.19437},
  year={2024}
}

@article{yang2024qwen2,
  title={Qwen2 Technical Report},
  author={Yang, An and Yang, Baosong and Hui, Binyuan and Zheng, Bo and Yu, Bowen and Zhou, Chang and Li, Chengpeng and Li, Chengyuan and Liu, Dayiheng and Huang, Fei and others},
  journal={arXiv preprint arXiv:2407.10671},
  year={2024}
}

@inproceedings{gale2023megablocks,
  title={MegaBlocks: Efficient Sparse Training with Mixture-of-Experts},
  author={Gale, Trevor and Narayanan, Deepak and Young, Cliff and Zaharia, Matei},
  booktitle={Proceedings of Machine Learning and Systems},
  volume={5},
  year={2023}
}

@inproceedings{hwang2023tutel,
  title={Tutel: Adaptive Mixture-of-Experts at Scale},
  author={Hwang, Changho and Cui, Wei and Xiong, Yifan and Yang, Ziyue and Liu, Ze and Hu, Han and Wang, Zilong and Salas, Rafael and Jose, Jithin and Ram, Parimarjan and others},
  booktitle={Proceedings of Machine Learning and Systems},
  volume={5},
  year={2023}
}

@inproceedings{he2022fastermoe,
  title={FasterMoE: Modeling and Optimizing Training of Large-Scale Dynamic Pre-Trained Models},
  author={He, Jiaao and Zhai, Jidong and Antunes, Tiago and Wang, Haojie and Luo, Fuwen and Shi, Shangfeng and Li, Qin},
  booktitle={Proceedings of the 27th ACM SIGPLAN Symposium on Principles and Practice of Parallel Programming},
  pages={120--134},
  year={2022}
}

@article{tan2024scattermoe,
  title={ScatterMoE: Efficient Mixture-of-Experts with Scatter and Gather Operations},
  author={Tan, Shawn and Shen, Yikang and Chen, Zhenfang and Courville, Aaron and Gan, Chuang},
  journal={arXiv preprint arXiv:2403.08245},
  year={2024}
}

@inproceedings{tillet2019triton,
  title={Triton: An Intermediate Language and Compiler for Tiled Neural Network Computations},
  author={Tillet, Philippe and Kung, H. T. and Cox, David},
  booktitle={Proceedings of the 3rd ACM SIGPLAN International Workshop on Machine Learning and Programming Languages},
  pages={10--19},
  year={2019}
}

@inproceedings{kwon2023vllm,
  title={Efficient Memory Management for Large Language Model Serving with {PagedAttention}},
  author={Kwon, Woosuk and Li, Zhuohan and Zhuang, Siyuan and Sheng, Ying and Zheng, Lianmin and Yu, Cody Hao and Gonzalez, Joseph and Zhang, Hao and Stoica, Ion},
  booktitle={Proceedings of the 29th Symposium on Operating Systems Principles},
  pages={611--626},
  year={2023}
}

@inproceedings{lepikhin2021gshard,
  title={GShard: Scaling Giant Models with Conditional Computation and Automatic Sharding},
  author={Lepikhin, Dmitry and Lee, HyoukJoong and Xu, Yuanzhong and Chen, Dehao and Firat, Orhan and Huang, Yanping and Krikun, Maxim and Shazeer, Noam and Chen, Zhifeng},
  booktitle={International Conference on Learning Representations},
  year={2021}
}

@misc{amd2024triton,
  title={Triton Support for AMD Instinct GPUs},
  author={{AMD}},
  howpublished={\url{https://rocm.docs.amd.com/projects/triton/en/latest/}},
  year={2024}
}

\end{document}